\documentclass[conference]{IEEEtran}
\IEEEoverridecommandlockouts
\usepackage{cite}
\usepackage{amsmath,amssymb,amsfonts}
\usepackage{algorithmic}
\usepackage{graphicx}
\usepackage{textcomp}
\usepackage{xcolor}

\usepackage{hyperref}
\usepackage{color}
\usepackage[normalem]{ulem}

\usepackage[english]{babel}
\usepackage[utf8]{inputenc}
\usepackage{amsmath}
\usepackage{amsfonts}
\usepackage{graphicx}
\usepackage[colorinlistoftodos]{todonotes}
\usepackage{float}

\usepackage{authblk}
\usepackage{graphicx}
\usepackage{subcaption}

\usepackage[ruled]{algorithm2e}
\usepackage{float}
\DontPrintSemicolon

\usepackage{authblk}
\usepackage{graphicx}
\usepackage{subcaption}

\usepackage{enumitem}
\usepackage{balance}
\usepackage{csquotes}

\def\BibTeX{{\rm B\kern-.05em{\sc i\kern-.025em b}\kern-.08em
    T\kern-.1667em\lower.7ex\hbox{E}\kern-.125emX}}

\begin{document}

\title{3D Marketplace: Distributed Attestation of 3D Designs on Blockchain}

\author{Sofia Belikovetsky\qquad Oded Leiba \qquad Asaf Shabtai \qquad Yuval Elovici \\ Department of Software and Information Systems Engineering \\ Ben-Gurion university of the Negev}

\maketitle


\begin{figure*}[!h]
  \includegraphics[width=\textwidth,height=4cm]{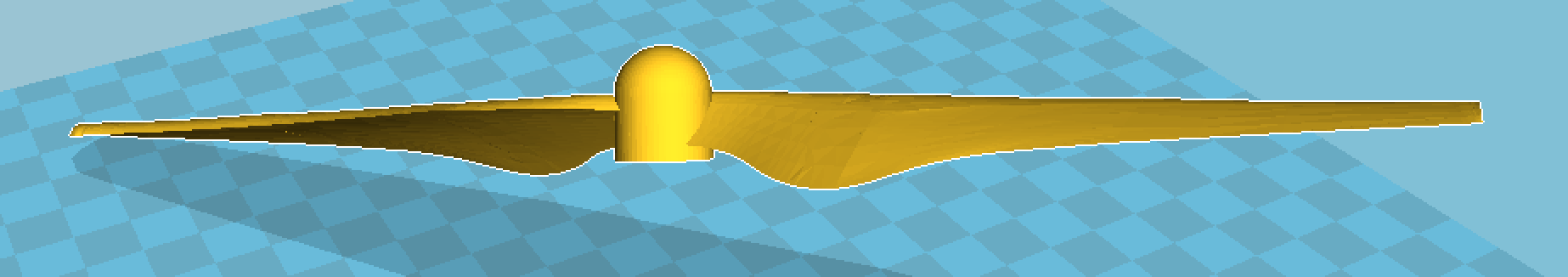}
  \caption{Faulty 3D design file}
  \label{fig:3dpart}
\end{figure*}

\begin{abstract}
Additive manufacturing (AM), or 3D printing, is an emerging manufacturing technology that is expected to have far-reaching socioeconomic, environmental, and geopolitical implications. 
As the use of this technology increases, the need for validation of 3D designs grows. 
In order to create a marketplace in which 3D designs are traded, 
it is necessary to develop a platform that enables the attestation of the 3D designs and promotes truthfulness. 
In this paper, we introduce a novel concept of a distributed marketplace that will support the attestation of 3D printing designs.
We build a mathematical trust model that ensures truthfulness among rational, selfish, and independent agents, which is based on a reward/penalty system. 
The payment for participating in the evaluation is calculated by factoring in agents' reputation and peer feedback. 
Moreover, we describe the architecture and the implementation of the trust model on blockchain using smart contracts for the creation of a distributed marketplace. 
Our model relies both on theoretical and practical best practices to create an unique platform that elicit effort and truthfulness from the participants. 
\end{abstract}


\section{Introduction}

Additive manufacturing (AM), which is often referred to as 3D printing, is a manufacturing technology that creates objects by incrementally fusing layers of material together.
This manufacturing technology can be applied to polymers, metals, alloys, and composites.
AM has numerous technological, environmental, and socioeconomic advantages, such as the ability to manufacture objects with complex internal structures, shorter design-to-production time, just-in-time and on-demand production, and reduced raw material waste.
These advantages encourage the use of AM to manufacture functional parts and prototypes for medical implants \cite{murr2010next}, for military usages like the recent 3D-printed grenade launcher \cite{hodkins2017meet}, and many critical cyber-physical systems.
An example from the aviation field is the FAA-approved fuel nozzle for General Electric's LEAP jet engine \cite{GE2015faa}.

AM's dependence on computerized systems has raised concerns in the security community regarding possible scenarios of IP theft and sabotage attacks \cite{yampolskiy2018security}.
The recent "dr0wned" study \cite{belikovetsky17dr0wned} demonstrated a full chain of attack applied on AM and discussed the ways a manipulation can be inserted into the AM process.
The authors sabotaged a 3D printed rotor of a quadcopter UAV, causing the rotor to break mid-flight and the quadcopter to fall after a short operating time.
This, and other scenarios that have been discussed in the literature, can be carried out through manipulation of the design files (e.g Figure \ref{fig:3dpart} that contains a propeller with a design flaw that cannot be easily detected).
As the AM field grows, the need for validated design files increases.
Currently, the design files for 3D printing are open-source and located in various publicly available repositories across the Web, where anyone can upload a design.
However, in order to create an ecosystem around the AM technology, there is a need for a marketplace of 3D design files that have been validated and deemed secure. Since evaluation of 3D design files require resources and it is hard to measure the amount of effort that was invested in the attestation, there is a need for a set of rules that will elicit the required effort. However, even if effort was invested, the participant might still choose a strategy of not reporting the truth. Thus, we need to develop a system that encourages truthfulness and penalizes falsehood.

In this paper, we present a framework that facilitates a distributed marketplace for the validation and trade of 3D design files.
We present a model that enables users to test new designs and receive compensation for their efforts. 
Once the designs are validated, they can be sold in the proposed marketplace.
We build a mathematical model to create a platform that encourages truthfulness and is based on a reward/penalty system that takes reputation and peer feedback into account.
The effectiveness of this distributed marketplace depends heavily on the implementation of this trust model.
Therefore, the platform is implemented on top of the blockchain technology and guarantees complete transparency, ensuring that any design file that is evaluated is coupled with the voting history of all of the users, and the calculated result is publicly verifiable by anyone.

The main contributions of this paper are:
\begin{itemize}
\item We present a unique, distributed marketplace that allows 3D printing designers to distribute and sell their designs. Each 3D design is verified by the community and assigned with an integrity score;
\item The proposed framework is based on a novel reward/penalty system that takes into account reputation and peer feedback;
\item The proposed reward/penalty-based system is generic and can support other use cases such as a marketplace for open source code distribution and bug reporting system.
\end{itemize}

The remainder of the paper is structured as follows.
After discussing related work in section \ref{sec:relatedWork}, we present the threat model and motivation in section \ref{sec:threatModel}, and list the preliminaries in section \ref{sec:preliminaries}. 
Sections \ref{sec:trustModel} and \ref{sec:blockchain} describe the trust model and blockchain implementation of the flow for validating a design file, respectively.
In section \ref{sec:discussion} we discuss the model, possible real-world fraud scenarios and future work.
\section{Related Work}
\label{sec:relatedWork}

By the end of 2017, approximately 70 publications addressed the threats of AM security. 
A survey by Yampolskiy et al. \cite{yampolskiy2018security} discussed the ways in which the 3D design files can be manipulated. 

In the dr0wned study \cite{belikovetsky17dr0wned} researchers presented a full chain of attack with AM that caused material fatigue of a functional part by manipulating the 3D design files. 
The authors sabotaged the 3D printed rotor of a quadcopter UAV, causing the rotor to break and the quadcopter to fall from the sky after a short period of flight. 
In \cite{sturm2014cyber} the authors demonstrated that a part's tensile strength can be degraded by introducing defects such as voids (internal cavities) into the 3D design files.

Currently, there is no efficient way to determine whether a 3D design will produce a well-functioning 3D object. 
Mechanical engineers rely on computerized simulations to reduce the chance of a faulty design. 
In \cite{sinha2001modeling} Rajarishi Sinha et al. presented an overview of modeling and simulation technologies that are used to validate 3D designs. 
In their paper, they discuss both the tools for simulations and the limits of those methods. 

To create the marketplace that is presented in this paper, we rely on previous work in several fields. 
The fundamental trust aspects of the marketplace are inspired by the common e-commerce trust models \cite{corbitt2003trust}; we included the concept of reputation in our marketplace. Our formulas were inspired by the basic model of reputation for e-commerce which was introduced by Xiong and Liu in \cite{xiong2004peertrust}.

The mathematical model is based on game theory models of output agreement mechanisms that pay agents that agree with their peers when performing tasks that require effort.
In \cite{dasgupta2013crowdsourced} the authors present a model that elicits effort and truth telling of the participating agents, but the model assumes prescreening of agents. 
The work of Witkowski et al. \cite{witkowski2013dwelling} builds upon the research done in \cite{dasgupta2013crowdsourced} and relaxes the screening constraint and proves that truthfulness and effort can be elicited through determining the values of rewards and penalties. 

The need for a distributed market for 3D designs for AM has been addressed in industry white papers; for example, in \cite{adkins2018protecting} the authors discuss potential uses for blockchain technology for AM. 
The authors present a high level architecture of how decentralized and distributed manufacturing platform should be constructed. 
Thus far, the majority of work on blockchain for AM has focused on copyrights preservation \cite{holland2017copyright} and securing the supply chain \cite{kennedy2017enhanced}.

The blockchain implementation is built on top of Ethereum \cite{buterin2014next} and consists of smart contracts. An  example of creating a marketplace on top of blockchain was presented by Chlu in \cite{chlu2018reviews}. In the white paper, the authors present a platform for e-commerce that enables online shops to integrate their payments and reputation systems into the platform. 
Recently, many research papers on integration of Blockcahin in IoT were published. Christidis et al. reviewed Blockchain and smart contract applications for IoT in ~/cite{christidis2016blockchains} that also discusses the facilitation of the sharing of services and resources between devices. 
In ~\cite{miller2018blockchain}, the authors present a platform for Industrial Internet of Things based on
the Blockchain technology. The platform enables peers in a trustless network to interact with each other without the need for a trusted intermediary.

\section{The Threat Model and Motivation}
\label{sec:threatModel}

Typically, the 3D printing process starts with the 3D design file, which is the blueprint of the final 3D object. The format of the design files is generally CAD that is transformed to STL, AMF or 3MF file formats. Minor changes to the design file in this phase can completely alter the printed result \cite{belikovetsky17dr0wned}. \\
Before printing, the design has to be \enquote{sliced} into individual layers and translated to 3D printed instructions that can be sent to the 3D printer. Open source software, such as \emph{Slic3r} and \emph{Cura}, is commonly used for desktop 3D printers that employ FDM technology, and the resulting file, containing the instruction set, is produced. The file describes the tool path that is sent to the 3D printer via USB, SD card or network connection. The tool path is commonly composed of \emph{G-Code} commands, a legacy language for CNC machines.
A change to the tool path file can have the same harmful effects as a change to the original design files. However, a malicious modification in this phase is harder to detect since there is no easy way to reverse the G-code command back to the original design. 
Regardless of the compromised representation, 
the \emph{cyber-phy\-si\-cal} impact depends on the physical change made to the printed object. \\
Predictions about the future of AM describe scenarios where anyone can download design files, customize them, and manufacture anything \cite{mota2011rise}. Currently, there are two categories of AM users. 
The first category consists of home users or hobbyists that manufacture 3D objects in a home environment, mainly using fused filament fabrication (FDM) technologies.
Design files are either self-made or downloaded from publicly available repositories. 
In this case, the harmful effect of a malicious or flawed design is usually limited; it can result in wasted time or materials, but only in extreme cases cause harm.
The second category consists of industrial users who design and manufacture 3D objects and functioning parts for industries, such as the automotive and aviation industry, or for military use. 
The design files are created either in-house or imported via a trusted supply chain. However, in both use cases, the design files are the weakest link in the manufacturing process. Thus, there is a need in adding a security layer for the validation of 3D design files. However, validation of design files is a complex task that requires expertise and cannot be easily automated. To solve this problem, we suggest constructing a marketplace that encourages the creation and sale of such designs.
Since AM technology is fragmented across many sectors, and there is currently no entity that can ensure quality testing across all of the different disciplines. 
Moreover, even if such an entity existed, issues of trust with rival companies and the threat of closed source logic for calculations of reputation and rewards would remain a problem.
Therefore, we cannot rely on a centralized marketplace, there is a need for a decentralized solution, where designs are decoupled from their creators and tested by the community, and only then are sold. 
In order to ensure transparency, we use Blockchain. 
It enables transparency of both the vote history of the users and the code that is used to calculate the results. 
All of the data needed for calculation, especially payment calculations, is stored and publicly verifiable. 
This guarantees that no censorship takes place by a single entity and the results are not biased. 


%

\section{Preliminaries}
\label{sec:preliminaries}

\subsection{Transaction Definition}

We define a transaction process $G$ to be the entire process a 3D design undergoes in the trust model. 
Let $G = <d, EP, FP, \tau_r, \tau_p, q^*>$ where,  
\begin{itemize}
\item $d$ is the design file. The design file also contains the printing instructions that specify how to translate the design to the 3D printer instructions set;
\item $EP$ (evaluation players) and $FP$ (feedback players) are sets of players in the evaluation phase and feedback phase, respectively. A player can participate either in the evaluation phase or the feedback phase $ EP \cap FP = \emptyset$. Since the rewards and penalties of $EP$ are determined based on the feedback of $FP$, thus these group have to be distinct to promote truthfulness;
\item $\tau_r$ and $\tau_p$ represent the evaluation phase reward and penalty, respectively;
\item the desired reputation level/quality level of the result is denoted by $q^* > \frac{1}{2}$.
\end{itemize}
We use $T(i)$ to denote the set of transactions in which player $i$ has participated and $D(j)$ to denote the set of players in transaction $j$. 

\subsection{Transaction Output}

The output of transaction process $j$ is a tuple $output_j = <r_j, FS_f(j)>$ where,
\begin{itemize}
\item $r_j \in \{-1, 0, 1\}$ is the result that indicates whether the design is valid ($1$ denotes a valid design, $-1$ denotes an invalid design, and $0$ indicates that the validation of the design could not be determined);
\item $FS_f \in [0,1]$ indicates the normalized final score after the feedback phase and can refer to the correctness of $r_j$ (also referred to as "weighted majority voting").
\end{itemize}
Each player $i$ in $G$ has a reputation $rep(i) \in [0,1]$ which was gathered from $i$'s participation in previous transactions in the marketplace.
In this paper, we use $FS$ and $D$ notations to refer to general concepts that apply to both phases (evaluation and feedback).


We distinguish between the final scores of each phase. We use $FS_e$ ($D(j) = EP$) and $FS_f$ ($D(j) = FP$) to denote the weighted majority voting of the evaluation phase and the feedback phase respectively.

\section{The Trust Model}
\label{sec:trustModel}

In this section we describe the stages of the trust model. 
The trust model starts with a 3D design file that is uploaded for evaluation and ends with the 3D design file being sold in the marketplace or taken down due to inadequate quality.
In this section we describe the process, providing a high level overview and a description of each phase of the process. 

\subsection{High Level Process Flow}
Similar to traditional e-commerce models, the purpose of the marketplace is to sell goods; specifically, to sell 3D designs to additive manufacturers. 
However, unlike traditional settings, general feedback about the quality of the design is insufficient. 

Validation of a 3D design can be done in various ways and levels of reliability. The testing can be done by design experts that validate the 3D design properties and best-practices, or it can be done in a more practical method by printing out the design and testing the 3D object. In this paper, we focus on validation that involves comprehensive real world testing that is both expensive and time consuming. 
The design needs to be printed, as well as potentially integrated as a functioning part in a different system or object, and the system or the object need to be tested.

\begin{figure*}
	\centering
	\includegraphics[width=.6\textwidth,height=.4\textheight]{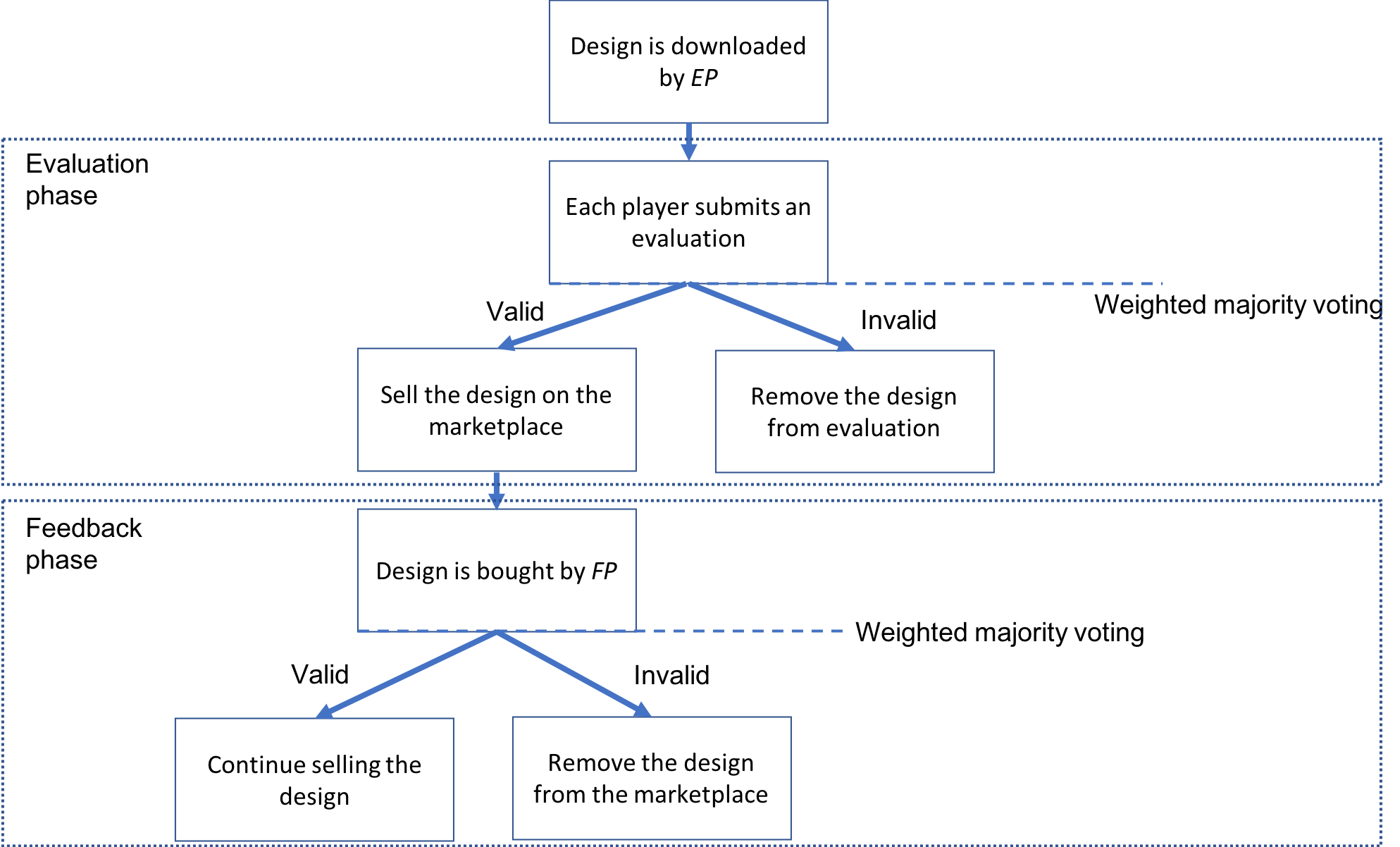}
	\caption{Diagram of the trust model}
	\label{fig:trust_model}
\end{figure*}

To incentivize users to perform such evaluation, we provide financial rewards for participation in the evaluation phase. 
To prevent abuse, financial penalties may be applied if users are not truthful or disagree with others. Our two phase trust model is presented in Figure \ref{fig:trust_model}. 

Before a design is permanently sold in the marketplace, it must go through the phases of the trust model.
In the first phase, the \enquote{evaluation phase,} the design is downloaded for free and evaluated by the $EP$. 
If, as a result of the vote, the design is determined to be valid, it is sold in the marketplace. 
The second phase, the \enquote{feedback phase,} depends on gathering feedback from the $FP$ (who purchased the design), in order to act as a \enquote{fail safe} against errors that may have occurred in the first phase. 
If the design passes the quality threshold $q^*$ in both phases, it is sold in the marketplace. 
Otherwise, it is taken down. 

\subsection{Reputation}

\subsubsection{Reputation Calculation}
\label{sub:RepCalc}

Each player $i$ in the marketplace has a reputation $rep(i)$ which is calculated based on the final score $FS(j)$ for every $j \in T(i)$. 
To define reputation in the proposed trust model, we adapt the basic model of trustfulness discussed in \cite{xiong2004peertrust} for calculation of reputation. In their work, Xiong et al. construct a well known model that computes trustfulness of peers by accounting basic trust parameters like feedback from other peers, the total number of transactions that the peer performs and credibility of the feedback sources.
In our model, the reputation value of each player is calculated based on a comparison of the player’s votes in previous transactions and the final scores of these transactions, reflecting whether the player voted along with the weighted majority.  
We define $a_i(j)$ as the answer (or vote) of player $i$ in transaction $j \in T(i)$ in the following manner: 

\begin{equation}
a_i(j) = \begin{cases} 
	1, & \mbox{if } d\mbox{ is valid} \\
	-1, & \mbox{if } d\mbox{ is invalid} \\
	0,  & \mbox{no answer or can not decide} 
\end{cases}
\end{equation}
 
The vote can apply to the evaluation phase or the feedback phase, depending in which phase player $i$ participated in transaction $j$. 
Then we define $rep(i)$ as: 

\begin{equation}
\label{eq:reputation}
rep(i) = \frac{1}{2} \cdot (\frac{\sum_{j \in T(i)} a_i(j) \cdot r_j \cdot FS(j)}{\sum_{j \in T(i)} FS(j)} + 1)
\end{equation}

The main component of equation \ref{eq:reputation} is terminated by the answer $a_i(j)$ that the player $i$ provided in transaction $j$ multiplied by the final score of that transaction. 
Thus, transactions with a higher final score will have greater impact on the reputation of agent $i$. 
The value $r_j$ indicates the trajectory of the scoring, $r_j = 1$ means that the design is valid with the collective vote of $FS(j)$, and $r_j = -1$ means that the design is invalid with the collective score of $FS(j)$. 
When combined with the value $a_i(j)$, $rep_i(j)$ increases (the reputation builds) for each transaction $j \in T(i)$ where $sign(a_i(j)) = sign(r_j)$. Thus, $rep_i(j)$ increases only when the player agreed with the weighted majority in transaction $j$. 
Otherwise, the reputation decreases, meaning that the value of $rep_i(j)$ decreases and lowering the impact player $i$ has on future transactions.  
The value of $FS(j)$ influences the amount the reputation increases or decreases thus the level of future influence of player $i$.
Finally, we normalize the equation.

\subsubsection{Weighted Majority Voting}

The weighted majority voting (or final score) is calculated in both phases of the trust model in the same manner.
The final score of transaction $j$ is calculated based on the answers $a_i$ for each player $i \in D(j)$ and their reputation $rep(i)$. 
Higher reputation values have greater impact on the final score of the transaction. However, the reputation of a player cannot stand alone in this setting, since the amount of transactions the player has should also be a factor, e.g., player $i_1$ with reputation $R$ who participated in $\vert T(i_1) \vert = 10$ transactions should have less influence on the final score than player $i_2$ with reputation $R$ that participated in $ \vert T(i_2) \vert = 100$ transactions, specifically, 10 times less influence.
Therefore, we add a component of weight to the reputation of each player. 

\begin{equation}
w_i(j) = \frac{\vert T(i) \vert}{\sum_{k \in D(j)} \vert T(k) \vert} 
\end{equation}

We calculate the final score $FS$ as follows: 

\begin{equation}
\label{eq:fs}
FS(j) = \frac{1}{2} \cdot (\frac{\sum_{i \in D(j)} a_i(j) \cdot rep(i) \cdot w_i(j)}{\sum_{i \in D(j)} rep(i) \cdot w_i(j)} + 1)
\end{equation}

The value of equation \ref{eq:fs} is mainly determined by the scalar product of  $a_i(j)$ 
$\forall i \in D(j)$ and the weighted reputations of the players $rep(i) \cdot w_i$. 
Thus, the votes of players with a higher weighted reputation have greater impact on the final score. 
Since $a_i$ is a signed value and the main component of the equation derives a value $\in [-1,1]$; we normalize this value for the final score. 

The results of a weighted majority vote are as follows: 
\begin{equation}
\begin{cases} 
r_j = 1 , & \mbox{if } FS(j) > q^*\\
r_j = -1 , & \mbox{if } FS(j) < 1 - q^* \\
r_j = 0,  & \mbox{otherwise}
\end{cases}
\end{equation}

If the result of the weighted voting is larger than $q^*$,  the design is determined to be valid. 
If $FS(j) < 1 - q^*$, the design is determined to be invalid. 
Otherwise, the game is annulled, and $r_j = 0$.

\subsection{Compensation Mechanisms}

\subsubsection{Rewards and Penalties} 

\begin{figure}
	\centering
	\includegraphics[width=.5\textwidth,height=.15\textheight]{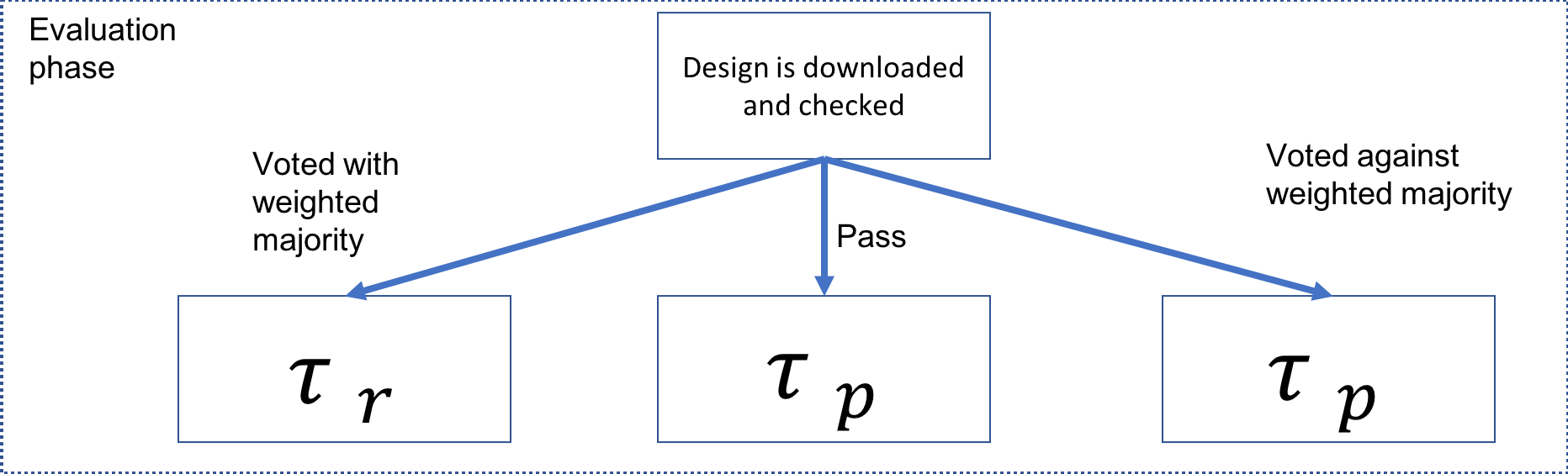}
	\caption{Diagram of the rewards and penalties}
	\label{fig:rewards}
\end{figure}

In the evaluation phase, the design is uploaded to the marketplace and available for free to any tester $i \in EP$. 
Every tester $i$ must provide an answer $a_i^e$ at the end of the evaluation phase, where $a_i^e(j) = a_i(j)$ $\forall j \in EP$. 
The reward or penalty for each player is determined both by the weighted majority and, in retrospect, by the voting in the feedback phase.  
Figure \ref{fig:rewards} presents the rewards and penalties a player may receive during a transaction. 
In the evaluation phase, since every participating player must provide a rating, players who do not answer, choose to pass after downloading the design, or do not reveal their vote as discussed in section \ref{sec:blockchain}, will pay a penalty of $\tau_p$. 
The payment for other players, who have provided a valid/invalid rating, is determined by the weighted majority of their peers. 
If the weighted majority has agreed with them, they get a reward of $\tau_r$ and a penalty of $\tau_p$ otherwise. 
We discuss how to assign specific values to the rewards and penalties in section \ref{sub:truthfulness}.


\subsubsection{Feedback}

The reputation of a player who has voted with the weighted majority will increase in both phases. 
Equation \ref{eq:reputation} describes the calculation of the reputation for each player. 
For each vote that corresponds with the weighted majority in transaction $j$, the reputation will increase by $FS(j)$ and be normalized. 
Thus, in the evaluation phase, players that vote with the weighted majority receive a reward of $\tau_r$, and their reputation grows; players who voted against the weighted majority receive a penalty of $\tau_p$ and their reputation decreases. 
Since players already know that the design has been rated high enough to be sold, there may be confirmation bias in the feedback phase, such that $FP$ players might agree with the valid rating more easily.
Thus, the model does not provide financial compensation for the feedback players. 
However, lack of motivation might result in biased feedback, 
and in order to prevent that, the reputation of players who voted with the weighted majority will increase. Similarly, the reputation of players that voted against the weighted majority will decrease.

\subsection{Truthfulness}
\label{sub:truthfulness}

We assume that the players in each transaction are rational, selfish, and independent agents. 
The output agreement presented in this section describes the way in which a player receives a reward for agreeing with his/her peers and a penalty for disagreeing. The desired strategy that leads to Nash Equilibrium in the proposed model is when the players answer truthfully after investing effort in testing the 3D design. Smartly choosing the values of the compensations will compel the players to choose this strategy. 
We rely on the results of Witkowski et al. that were presented in \cite{witkowski2013dwelling} which showed that with the appropriate payments, agents prefer passing (not participating) to guessing, and moreover that if they choose to participate and invest effort, truthfulness is their best strategy. 
We logically reduce their problem to the problem presented in this paper and apply their findings to the proposed model in order to determine the suitable compensation values that will lead to the desired Nash Equilibrium. 
In the paper, the authors describe a game in which there are two items $A$ and $B$ that need ranking. 
Each selfish and independent agent has to invest effort to observe which of the items is better. 
Agents get a reward for agreeing with their peers and a penalty for disagreeing. 
We rely on their results and adapt them to our model in the following way. 

In our model, without the loss of generality, let $A$ be a valid design and $B$ be an invalid design. 
Let $C^*>0$ be the cost of effort. 
We define $C^*$ as the cost of the material that is needed to 3D print the design. 
Since the amount of material can be derived from the design and metadata of the design, it can be estimated and will be the same for all players that follow the printing instructions. 
The authors discuss a self-selecting mechanism for which every agent below a specific qualification factor will choose not to participate. 
The qualification factor indicates the probability of observing the correct answer after investing effort. 
In our model, we defined this minimal quality factor as $q^* \in [\frac{1}{2}, 1]$. 
We define $x^*$ to be the normalized value of $q^*$, so that $q^* = \frac{x^* + 1}{2}$ and $x^* = 2 \cdot q^* -1$. 
In \cite{witkowski2013dwelling} the authors prove that a certain amounts of payments induce truth-telling in the describes settings. We project their model onto our model and define the payments in the same way. 
Thus, the payment for truth-telling should be: 

$$ \tau_r = \frac{2C^*}{(x^*)^2 + x^*}$$

\begin{equation}
\label{eq:reward}
 \tau_r  = \frac{C^*}{2 \cdot (q^*)^2}
\end{equation}

and the penalty payment should be: 

$$  \tau_p = -\frac{2C^*}{(x^*)^2 + x^*} - \epsilon $$

\begin{equation}
\label{eq:penalty}
\tau_p = -\frac{C^*}{2 \cdot (q^*)^2} - \epsilon
\end{equation}

for $\epsilon \rightarrow 0$.

Since our scoring method is more complex than the binary answer that is discussed in \cite{witkowski2013dwelling}, we explain how the payment for a player is determined. 
In the original work, each agent's answer is compared with a random peer, and if their votes match, the agent received a reward (or a penalty if their votes do not match). 
In our model, for each player $k$ in transaction $j$, we compare two independent values. We compare the player's weighted vote: $$sign(\frac{a_k(j) \cdot rep(k) \cdot w_k(j)}{\sum_{i \in D(j)} rep(i)})$$ with the weighted vote of the rest of the players:
$$sign(\frac{\sum_{i \in D(j) \setminus k} a_i(j) \cdot rep(i) \cdot w_i(j)}{\sum_{i \in D(j)} rep(i)})$$
if the signs match, meaning that the player voted along the weighted majority, the player gets a reward $\tau_r$; if otherwise the player receives a penalty $\tau_p$.  
For each player, we reduce the complexity by assuming that the player's vote is compared to another peer with the weighted answer and the reputation of the rest of the players. 
\section{Blockchain Implementation}
\label{sec:blockchain}

\begin{figure*}
	\centering
	\includegraphics[width=.8\textwidth,height=.5
    \textheight]{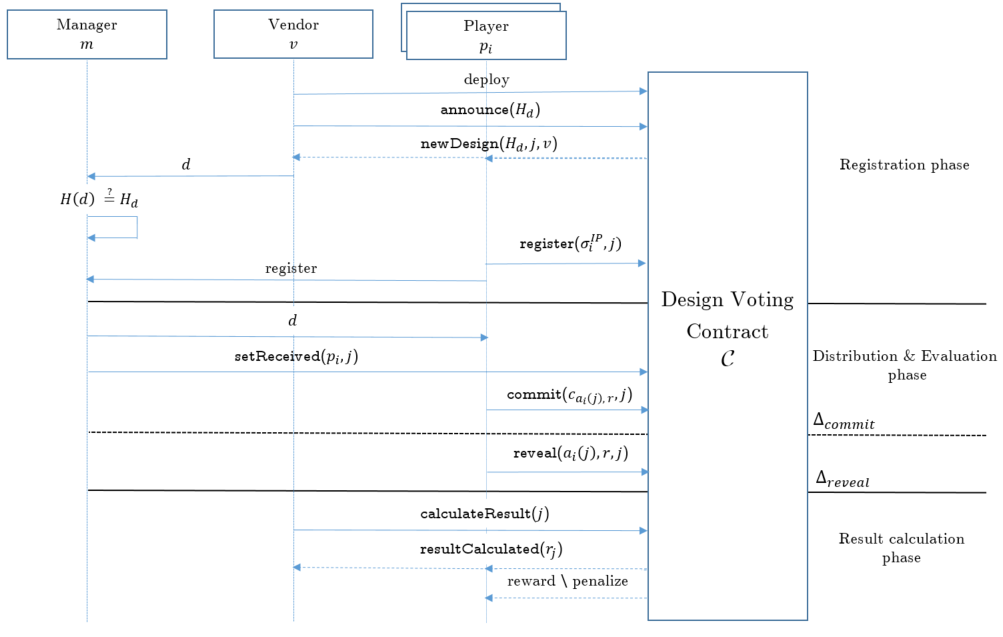}
	\caption{Outline of the entire design evaluation process using a design voting contract}
	\label{fig:bcGeneral}
\end{figure*}

In this section, we describe the blockchain implementation and procedure sequence of the distributed attestation.
A design voting smart contract is constructed to execute the trust model rules as a result of the messages (blockchain transactions) it gets from the entities (accounts) in the system. 
Figure \ref{fig:bcGeneral} shows the sequence of the procedures and the interaction between the parties involved: the evaluating players, the vendor, the design voting smart contract, and a semi-trusted \textit{manager}. 
The process consists of three phases: the registration phase, distribution and evaluation phase, and finally, the result calculation phase.\\
In section \ref{involved} we describe the parties involved in more depth, and in section \ref{protocol} we provide a description of the protocol and the different steps in a distributed attestation of a single design.

\subsection{\label{involved} Parties Involved}
\begin{description}
\item \textbf{Vendor.} A vendor, i.e. the designer of a 3D design is denoted by $v$. 
The vendor’s interest in the framework is to obtain a transparent and reliable evaluation of the proposed design, so a large pool of potential buyers can be convinced of the designs validity and be encouraged to download and use it.\par
\item \textbf{Players.} A player, i.e., a single consumer from the set $EP$ (or $FP$ when in feedback phase), is denoted by $p_i$. A third party identity provider, denoted by $IP$, is required to sign the public address of $p_i$ in the blockchain to attest that the player is linked to a unique authenticated real-world identity.\par
\item \textbf{Design Voting Smart Contract.} The smart contract, denoted by $\mathcal{C}$, is responsible for the publicly verifiable calculation of the weighted majority voting for each design, with respect to the formulas presented in section \ref{sec:trustModel}. 
The contract stores the players' states: their reputation, voting weight, and previous votes. 
Evaluation$\backslash$feedback messages are sent by the players to the contract, and eventually, a design's $<r_j, FS(j)>$ is calculated using the participating player's reputations and recent votes. 
The pseudocode for the smart contract is presented in Algorithms \ref{alg:contractpseudo} and \ref{alg:contractpseudo2}. \par
\item \textbf{Manager.} A semi-trusted entity, denoted by $m$, ensures a fair exchange of a 3D design for an evaluation. Specifically, the manager is trusted for ensuring that player $p_i$ is marked as a receiver of the design file in the system \textit{if and only if} he/she indeed received the design. 
This is needed for a trustworthy execution of the design voting contract. 
Note that the manager is not trusted to ensure the reliability of the evaluation messages of the players. Also, the integrity of the messages that are handled by the manager can be checked since a player registration is verified against his/her public address and a corresponding signature by an identity provider on that same address.
Additionally, the manager is not involved in the calculation of the final score and the rewarding and penalizing of the players, as these are done in a publicly verifiable way by the smart contract.\\
To minimize the dependence in the manager further, the manager may be implemented by a consortium of authorities to distribute the trust, or by using a trusted execution environment such as Intel SGX. 
These options are out of the scope of this paper and left as a direction for future work.\\
The public key of $m$ is hard-coded into the smart contract $\mathcal{C}$, and all participating parties must create an reliable and authenticated channel using $m$. (i.e. authenticates messages passed in it with respect to the public key of $m$ and the public key of the counter party as was published in the smart contract, and every packet sender knows whether it reached its destination). 
\end{description}

\subsection{\label{protocol} Protocol Description}

In this section we describe the sequence of procedures performed as part of the distributed attestation protocol.
The flow consists of three main phases: the registration phase, distribution and evaluation phase, and result calculation phase.\par
\begin{enumerate}
\item{\textbf{Registration Phase:}} First, some party deploys the design voting smart contract $\mathcal{C}$.
The deploying party should not necessarily be trusted, as the contract code is meant to be published and can be independently audited by anyone. 
Without loss of generality, we say that $v$ itself deploys the contract. 
The contract is meant to host the logics for the voting and evaluation of \textit{all} designs, hence a setup step of registration needs to take place only once.\\
Second, a vendor $v$ announces a new available design $d$ by sending \texttt{announce(}$H_d$\texttt{)} to the smart contract, where $H_d$ is the design file identifier (a cryptographic hash digest of the file). The pseudo code of the command is detailed in algorithm \ref{alg:contractpseudo}.
Also, this message should contain a collateral from the vendor. This collateral will be used to later reward the participating players according to their voting.\\
As a consequence, the contract emits \texttt{newDesign(}$H_d, j, v$\texttt{)} as an event, where $j$ is the index of the design $d$ (transaction $j$).
The vendor and all potential players who listen to the event receive this notification.\\
Next, the semi-trusted manager $m$ is needed to conduct the fair exchange of the design $d$ for an evaluation. Specifically:\\
- $v$ sends $d$ to $m$.\\
- $m$ verifies that $H(d) = H_d$, where $H$ is the cryptographic hash function agreed upon in the protocol. 
If the condition does not hold, $m$ stops interaction with $v$ and does not continue distribution of this particular design.\\
- Each interested player $p_i \in EF$ (or $FP$ when in feedback phase) registers to the contract by sending a transaction to the function \texttt{register} (algorithm \ref{alg:contractpseudo}) with:
\begin{enumerate}
\item a signature of the identity provider $IP$ on his/her public address, denoted by $\sigma^{IP}_i$;
\item collateral - a deposit of at least $\tau_p$.
\end{enumerate}
The contract will verify the signature of $IP$ on the public address of $p_i$ and store the deposited collateral amount with an allocated initial empty state for $p_i$.\\
- $p_i$ is registered and authenticated via a private channel in front of the manager $m$ (i.e., proving possession of the private key associated with the public key he/she used for the registration in the contract).
\item{\textbf{Distribution and Evaluation Phase:}} This phase is spitted into three steps: 
(a) distribution step, (b) commit step, and (c) reveal step.
\begin{description}
\item{\textbf{Distribution step:}}\\
- $m$ sends $d$ to each $p_i \in EP$ (and each $p_i \in FP$ when in feedback phase).\\
- $p_i$ also checks $H(d) = H_d$.\\
- $m$ sends \texttt{setReceived(}$p_i, j$\texttt{)} to the smart contract. This call will set the respective flag reserved for $p_i$ and design $d$ (of index $j$) in the contract as true. This flag indicates that $p_i$'s vote should count. If $p_i$ does not vote, the player should be penalized.\par

\item{\textbf{Commit step:}} The \texttt{commit} step in the contract is a step where each player should send a cryptographic commitment of his/her vote with respect to a specific design announced.\\
 $\Delta_{commit}$ is a preset contract constant, which indicates the maximum time allowed between a vendor's \texttt{announce} message and a corresponding \texttt{commit} message by a registered player. Hence, the commit step for the design of index $j$ starts separately  for every player $p_i$ from the moment when $m$ has called \texttt{setReceived(}$p_i, j$\texttt{)} (algorithm \ref{alg:contractpseudo}), and ends collectively upon the expiration of $\Delta_{commit}$, after which all \texttt{commit} messages to that same design will be reverted (e.g., throw an error).\\
We are using a commit-reveal scheme with a shared commit phase end time per design, in order to prevent players from biased voting in which a player votes according to the votes of other players.\\
Therefore, in this step, $p_i$ sends \texttt{commit(}$C_{a_i(j), r}, j$\texttt{)} (algorithm \ref{alg:contractpseudo}), where $C_{a_i(j), r}$ is the cryptographic commitment to ${a_i(j)}$, using the blinding factor $r$.\par

\item{\textbf{Reveal step:}} The \texttt{reveal} step in the contract is a step in which each player should send the opening of his/her cryptographic commitment sent in the preceding \texttt{commit} message, and reveal the vote $a_i(j)$.\\
$\Delta_{reveal}$ is a preset contract constant, which indicates the maximum time allowed between $\Delta_{commit}$ and a \texttt{reveal} message with respect to the same design. Hence, the reveal step starts from the expiration of $\Delta_{commit}$ and ends with the expiration of $\Delta_{reveal}$, after which all sent \texttt{reveal} messages sent will be reverted (e.g., throw an error). Thus, in this step $p_i$ sends \texttt{reveal(}$a_i(j), r,  j$\texttt{)} (algorithm \ref{alg:contractpseudo}), where $r$ is the blinding factor used to create the commitment $C_{a_i(j), r}$. The state of $p_i$ is now updated in the contract with the player's vote.
\end{description}

\item{\textbf{Result Calculation Phase:}} Any party can initiate the result calculation phase by sending \texttt{calculateResult(}$j$\texttt{)} (algorithm \ref{alg:contractpseudo2}) to the smart contract $\mathcal{C}$. 
Since the outcome of the calculation phase is independent of the sender, any party can initiate the calculation.
Again, for brevity and without loss of generality, we say that a $v$ performs this step (if the $v$ is motivated not to send this message because calculating the result score may damage him/her, then any other player or external party can do this).
The smart contract, given the state of all of the players' current votes and calculated reputations from previous designs, calculates the design's final score $FS(j)$ and the respective result $r_j$ as described in the equations mentioned in section \ref{sec:trustModel}.\\ As a consequence, $\mathcal{C}$ emits \texttt{resultCalculated(}$r_j$\texttt{)} with the weighted majority voting result $r_j$ to all of the parties involved, and accordingly, all of the penalties or rewards can be collected or provided using the collateral of the parties involved (both $v$'s and the players' collateral). Penalties and rewards are also calculated in terms of individual reputation. Moreover, penalties are collected from players for which $m$ reported that the player received $d$, but did not provide an evaluation for $j$, in order to prevent players from \enquote{free-riding} and just getting the design for free, or alternatively, deciding not to reveal their vote in the reveal phase if they find out that they voted against the majority after viewing the other votes in clear text.
\end{enumerate}

\begin{algorithm}

\SetAlgoLined
\SetKwInOut{Input}{input}\SetKwInOut{Output}{output}
\SetKwInOut{contract}{contract}
\SetKwFunction{register}{register}
\SetKwFunction{announce}{announce}
\SetKwFunction{commit}{commit}
\SetKwFunction{reveal}{reveal}
\SetKwFunction{setReceived}{setReceived}
\SetKwFunction{calculateResult}{calculateResult}
    Initialize \textsf{playerStates} $:= \emptyset$\\
    Initialize \textsf{results} $:= \emptyset$\\
    Initialize \textsf{designStates} $:= \emptyset$\\
    Initialize constants: $\Delta_{commit}, \Delta_{reveal}, q, \tau_p, \tau_r, \epsilon, m$ \\
	\;
on \textbf{contract} \textbf{input} \announce$(H_d, \$X)$ from $v$ at time $T$: \\
\Indp $j := |$\textsf{designStates}$|$\\
	\textsf{designStates[}$j$\textsf{].vendor}$ := v$\\
	\textsf{designStates[}$j$\textsf{].T}$ := T$\\
    \textsf{designStates[}$j$\textsf{].balance}$ := \$X$\\
	emit \texttt{newDesign}$(H_d, j, v)$\\\Indm
    \;
on \textbf{contract} \textbf{input} \register$(\sigma^{IP}_i, j, \$X)$ from $p_i$: \\
\Indp discard if $\sigma^{IP}_i$ is an invalid signature of $IP$ on $p_i$\\
	discard if $\$X < \tau_p$ \\
    discard if the number of registered players for design $j$ together with $p_i$, exceeds \textsf{designStates[}$j$\textsf{].balance} $\div$ $\tau_r$\\
      \textsf{playerStates[}$p_i$\textsf{].reputation} $:= \epsilon$\\
      \textsf{playerStates[}$p_i$\textsf{].weight} $:= \epsilon$\\
      \textsf{playerStates[}$p_i$\textsf{].commitments} $:= \emptyset$\\
      \textsf{playerStates[}$p_i$\textsf{].votes} $:= \emptyset$\\
      \textsf{playerStates[}$p_i$\textsf{].received} $:= \emptyset$\\\Indm
	\;
on \textbf{contract} \textbf{input} \setReceived$(p_i, j)$ from $s$: \\
\Indp discard if $s \neq m$ \\
	\textsf{playerStates[}$p_i$\textsf{].received[}$j$\textsf{]} $=$ \texttt{TRUE}\\\Indm
	\;
on \textbf{contract} \textbf{input} \commit$(C_{a_i(j), r}, j)$ from $p_i$ at time $T$: \\
\Indp discard if \textsf{playerStates[}$j$\textsf{].received}  $\neq$ \texttt{TRUE}\\
	discard if $T >$ \textsf{designStates[}$j$\textsf{].T}  $+ \Delta_{commit}$\\
    \textsf{playerStates[}$p_i$\textsf{].commitments[}$j$\textsf{]} $:= C_{a_i(j), r}$\\\Indm
    \;
on \textbf{contract} \textbf{input} \reveal$(a_i(j), r, j)$ from $p_i$ at time $T$: \\
\Indp discard if $T >$ \textsf{designStates[}$j$\textsf{].T}  $+$ $\Delta_{commit}$ $+$ $\Delta_{reveal}$\\
    discard if $H(a_i(j), r) \neq$ \textsf{playerStates[}$p_i$\textsf{].commitments[}$j$\textsf{]}\\
    \textsf{playerStates[}$p_i$\textsf{].votes[}$j$\textsf{]}$ = a_i(j)$\\\Indm
    \;
 \caption{Design Voting Smart Contract - part 1}
 \label{alg:contractpseudo}
\end{algorithm} 

\begin{algorithm}

\SetAlgoLined
\SetKwFunction{calculateResult}{calculateResult}
on \textbf{contract} \textbf{input} \calculateResult$(j)$ at time $T$: \\
\Indp discard if $T \leq$ \textsf{designStates[}$j$\textsf{].T}  $+$ $\Delta_{commit}$ $+$ $\Delta_{reveal}$\\
    set $FS(j)$ as the calculation of the final score of transaction $j$ according to equation \ref{eq:fs}, using the players' reputation, weights and votes from \textsf{playerStates}, and the \textsf{results} array.\\
    set $r_j$ as the voting result using $FS(j), q$\\
    for each player $p_i$ s.t. \textsf{playerStates[}$p_i$\textsf{].votes[}$j$\textsf{]} $\neq$ \texttt{NULL}:\\
    \hspace{0.18cm}- send $p_i$ its deposit after increasing$\backslash$reducing a reward$\backslash$penalty according to the penalties$\backslash$rewards formulas. 
    \hspace{0.18cm}- set \textsf{playerStates[}$p_i$\textsf{].weight} and \textsf{playerStates[}$p_i$\textsf{].reputation} according to the reputation and weight formulas.\;
send \textsf{designStates[}$j$\textsf{].vendor} its remaining deposit\\
emit \texttt{resultCalculated}$(j, r_j)$\\\Indm
    \;
 \caption{Design Voting Smart Contract - part 2}
 \label{alg:contractpseudo2}
\end{algorithm}

\section{Discussion}
\label{sec:discussion}

In this paper we presented a model that is based on rational, selfish, and independent agents.
In real-world scenarios this may not be the case. 
There might be coalitions between agents that are working towards a common goal or \enquote{irrational} decisions that might result in financial loss but achieve some out of scope goal.
In this section, we try to suggest possible improvements to the model that can address those problems. 
We also discuss outstanding questions which will be addressed in future work. 

Since the model relies on trust, there are many opportunities for fraud. We try to address some of them here:
\begin{description}
	
\item Copyrights -  Since there is no correlation between the cost of the design and the penalty for not voting in the evaluation phase, if the design is costly, it might get stolen. 
A possible way to address this, is to watermark the design files so that if they are stolen, the player will be banned from the marketplace and penalized. 
Since the identity is derived from a third party entity, there might be a way to penalize the identity of the player in additional ways.
	
	\item Coalition -  The mathematical model that is used in this paper does not take into account possible collaboration between players. 
In order to detect such coalitions, an external mechanism can be constructed that aims at detecting players that vote together and restricting their participation. 
In future work we plan to address this issue and suggest such an external mechanism.
	
	\item Elicitation of Effort -  The trust model is built on the assumption that we cannot prove effort was invested. 
In AM there is a need to 3D print the design, possibly integrate it into an external system and test the system. 
We cannot provide a mathematical proof for the entire process, but future improvements in 3D printers might provide mathematical proofs for the 3D printing phase.
If a player can provide a proof-of-print (PoP), equation \ref{eq:fs} can be tweaked to give this player's vote more weight. 
One suggestion for how to implement this PoP is to add a trusted environment to 3D printers which would provide a mathematical proof for the activity of the 3D printer. 
Even though any PoP can be eventually hacked or forged, it can still act as an additional layer of security. 
In future work we will explore possible implementations for PoPs.

	 \item Gaining Reputation for a Single \enquote{Hit} - A player with external motivation might invest time and money in order to gain enough reputation to sway one or several transactions. 
     We believe that such a scenario cannot be addressed with logic or financial penalties. A possible way of reducing this risk is to rely on a strong third party identity provider. Another way is restricting the participation of players with a small number of previous transactions or with a low amount of gained reputation.
	
\end{description}

There are multiple points that we plan to address in future work:
\begin{enumerate}
\item We would like to extend the variety of players by allowing to validate a design in additional ways. In addition to physical testing, we can add validation by examining the soundness of the 3D design itself, or by running computer simulations on the design. Each player can choose the method of validation and each method will have different influence over the $FS$. Another parameter that we can integrate into the equation is the level of confidence the player has in his/her evaluation. 
\item In our blockchain implementation we rely on the smart contract keeping history of all the previous votes. 
Currently, this prompts non-trivial gas costs and in future work we will explore alternative storage options.
\item We plan to extend this model to additional use cases. This model excels in scenarios where there is a need in eliciting effort from experts. Thus, the proposed model can also be used to attest complex tasks like verification of open-source code by a community. This problem is at the heart of the trust in open-source software and a way to enhance the security and the reliance on the open-source code can have a great impact. Another use case is a bug reporting system, where each bug is tested by the community and voted upon.  
\end{enumerate}


In Conclusion, we believe that the proposed marketplace is an interesting and novel idea, since it suggests a way to attest a complex task with no known ground truth. 
The mathematical trust model elicits effort and truth-telling by utilizing a reward/penalty system. The payment for the players is calculated based on their votes and reputation compared to their peers' votes and the proposed implementation is built on blockchain which is decentralized and publicly verifiable infrastructure.



\balance
\bibliographystyle{IEEEtran}
\bibliography{bib}

\end{document}